\documentclass[aps,prx,reprint]{revtex4-2}

\usepackage{multirow}
\usepackage{soul}
\usepackage{nccmath}
\usepackage[export]{adjustbox}
\usepackage{pgfplotstable}
\usepackage{colortbl}
\usepackage{amsmath}
\usepackage{amssymb}

\DeclareMathOperator*{\argmax}{arg\,max}

\begin{document}

\title{Active Matter Flocking via Predictive Alignment}

\author{Julian Giraldo-Barreto and Viktor Holubec}
\affiliation{Department of Macromolecular Physics, Faculty of Mathematics and Physics, Charles University, 18000 Prague, Czech Republic}
\email{julian.giraldob@matfyz.cuni.cz, viktor.holubec@matfyz.cuni.cz}

\begin{abstract}
Understanding collective self-organization in active matter, such as bird flocks and fish schools, remains a grand challenge in physics. \textcolor{black}{Interactions that induce alignment are essential for flocking; however, alignment alone is generally insufficient to maintain group cohesion in the presence of noise, leading traditional models to introduce artificial boundaries or explicit attractive forces. Here, we propose a model that achieves cohesive flocking through purely alignment-based interactions by introducing predictive alignment, in which agents reorient to maximize alignment with the prevailing orientations of their anticipated future neighbors.}
 Implemented in a discrete-time Vicsek-type framework, this approach delivers robust, noise-resistant cohesion without additional parameters. In the stable regime, flock size scales linearly with interaction radius, remaining nearly immune to noise or propulsion speed, and the group coherently follows a leader under noise. These findings reveal how predictive strategies enhance self-organization, paving the way for a new class of active matter models blending physics and cognitive-like dynamics.
\end{abstract}

\maketitle

\emph{Introduction}:
From micrometer-sized bacteria to complex animals, biological organisms sense their environment, process directional cues, and adapt their motion accordingly~\cite{BERG1972, Ecology, AMRoadmap2025}. Similar feedback mechanisms are also indispensable in the control of autonomous robotic systems~\cite{Kaspar2021}. Based on visual~\cite{Pearce2014}, acoustic~\cite{Gorbonos2016}, or chemical~\cite{Robert2008} signals, these perception-reaction interactions result in the self-organization of large ensembles of cognitive individuals into cohesive spatiotemporal patterns, such as bird flocks~\cite{Ballerini2008}, fish schools~\cite{Berdahl2013}, and human crowds~\cite{Silverberg2013}. The study of these collective behaviors falls within the domain of active matter physics~\cite{Vicsek201271, BechingerReview2016, AMRoadmap2025}. Models of collective behavior in active matter span Reynolds-type ‘boid’ models~\cite{Reynolds1987}, Vicsek-type ‘alignment’ models~\cite{Vicsek1995,toner2024physics,Gompper2025}, Couzin-type ‘zonal’ models~\cite{Couzin2005}, ‘vision cone’ models~\cite{Barberis2016, LavergneBechingerScience2019}, motivation-based models~\cite{MehdiCrowds2011, RomanczukPursuitEscape2009, Charlesworth2019, Turner2023}, vision-based models~\cite{Pearce2014, Ito2024, Castro2025}, energy-efficiency models~\cite{Intesaaf2017}, and other biologically motivated models~\cite{Nagy2010}, as well as models designed for controlling robotic swarms~\cite{VicsekRobot2014}.

\textcolor{black}{Vicsek-type models that rely solely on alignment interactions struggle to maintain cohesion without artificial mechanisms such as periodic or reflecting boundaries, or additional attractive forces~\cite{Chate2008b, Solon2015b}. However, boundary conditions can influence bulk behavior, especially in the parameter regime associated with microphase separation, where density waves tend to align with the symmetries of the periodic simulation box~\cite{NAGY2007, Chate2008b, Solon2015b}.} Similarly, incorporating attractive interactions can induce swirling motion~\cite{Albaladejo2023}, which was absent in the original model. Other models achieve cohesion through either direct attractive interactions~\cite{Caprini2023} or explicit mechanisms, such as active or passive reorientation and movement toward a local or global center of the group~\cite{Reynolds1987, Couzin2005, Barberis2016, LavergneBechingerScience2019}. Notable exceptions include models where cohesion is not explicitly built into the algorithm, such as the maximum path entropy model~\cite{Charlesworth2019, Turner2023} or vision-based models~\cite{Pearce2014}. However, these approaches do not restrict the agents’ sight range, effectively introducing long-range interactions. To our knowledge, no prior model achieves cohesive flocking with purely alignment interactions over a finite range.

Here, we introduce predictive alignment in a Vicsek-type framework with a limited interaction radius \( \zeta \). We interpret the alignment interactions as biologically motivated social behaviors based on individual decision-making. \textcolor{black}{Specifically, we implement them using the sociological rule of “copy the other”~\cite{Rendell2010}, whereby an individual adopts the prevailing state of its neighbors—a strategy known to enhance individual success within a group.}

Our model reduces to a variation of the Vicsek model for simple agents that cannot anticipate future positions. \textcolor{black}{However, agents capable of anticipating their future neighbors effectively optimize a trade-off between alignment and proximity. This yields a cohesive flocking model based solely on alignment with the prevailing orientation of neighbors, without the need for additional parameters or boundary constraints.} The system undergoes a dynamical transition to an incoherent state with increasing noise and distance traveled per timestep over the interaction radius. \textcolor{black}{In the flocking state, the stationary flock radius is comparable to the interaction radius, independent of agent speed, and increases linearly with noise---albeit with a very small slope.} Additionally, the group efficiently follows a subgroup of maneuvering leaders. Our results reveal how predictive strategies enable robust self-organization akin to natural systems.

\emph{Model}:
\begin{figure}[ht!]
\centering
    \includegraphics[width=1.0\linewidth]{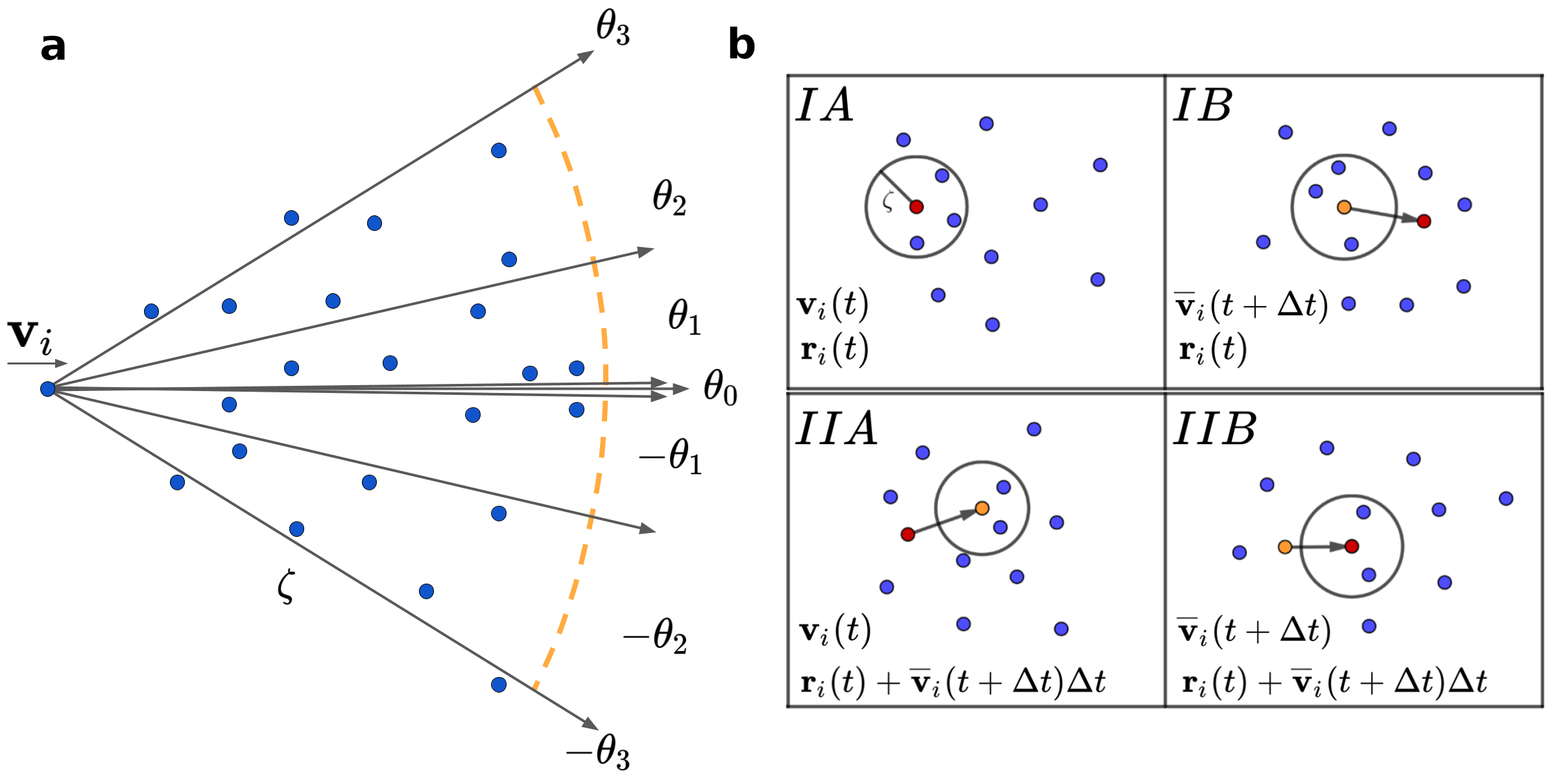}
   \caption{\textbf{Model.} \textbf{a} \textcolor{black}{At each discrete time step, individual agents aim to align as closely as possible with the prevailing orientation of their neighbors within a circle of radius $\zeta$.} To achieve this, they select one of seven possible reorientations, $\Delta \theta^t_i \in \Omega_\theta = \pm \{0, 0.01, 0.2, 0.5\}$, that maximizes the correlation function in Eq.~\eqref{eq:eq5}. All agents update their orientation in parallel. \textbf{b} We implemented four different strategies (IA–IIB) for evaluating the correlation function. In strategies I, the correlation is computed using the current neighbors ($\mathbf{r}_i = \mathbf{r}_i(t)$), whereas in strategies II, it is computed using predicted future neighbors [$\mathbf{r}_i = \mathbf{r}_i(t)+\overline{\mathbf{v}}_i(t+\Delta t)\Delta t$], as illustrated by the black circles. In strategies A, the agent's own orientation is included inside the bracket of the correlation function ($\mathbf{v}_i = \mathbf{v}_i(t)$), introducing orientational inertia, which is absent in strategies B ($\mathbf{v}_i = \overline{\mathbf{v}}_i(t+\Delta t)$).
}
    \label{fig:fig1}
\end{figure}
Biological active agents in nature follow evolutionarily adapted instincts and, in the case of higher animals, sometimes even learned or cognitively driven strategies to achieve specific goals such as collision avoidance or foraging.
 Similar mechanisms are also implemented in the development of autonomous robotic systems. These strategies are shaped by physical, biological, or technical constraints, which limit the range of possible dynamical and adaptive responses.
We consider a system of $N$ Vicsek-type agents self-propelling in discrete time in two dimensions with a constant velocity $v_0$ in the direction of their orientation vectors $(\cos\theta_{i}^{t}, \sin\theta_{i}^{t})$, $i =1,\dots,N$. At each discrete time step $t$, the agents reorient by discrete angles $\Delta \theta^t_i \in \Omega_\theta = \pm \{0, 0.01, 0.2, 0.5\}$ rads to achieve maximum alignment with their neighbors, as shown in Fig.~\ref{fig:fig1}a. \textcolor{black}{We used a discrete set of angles mainly for computational efficiency—selecting the optimal reorientation from a small, predefined set is significantly faster than solving the corresponding continuous optimization problem. The set $\Omega_\theta$ is chosen to allow agents to reorient either gradually or sharply, depending on how far their desired direction deviates from their current heading, to mimic the original Vicsek model. Nevertheless, as shown in Sec.~S8~\cite{supplementary}, a variant of the model with only three possible reorientation angles yields qualitatively similar results. The limited reorientation can be interpreted as a realistic constraint, reflecting the physical limitations of actual agents, such as friction or biomechanical restrictions that prevent abrupt turns. We also note that in the IIA and IIB variants of the model, discussed below and in Fig.~\ref{fig:fig1}b, the discrete angle sets effectively define agent’s field of view.}

\textcolor{black}{The imperfections in reorientation of the agents are reflected by a noise term $\xi_i^t$ sampled from the interval $\eta[-\pi, \pi]$, added to the chosen $\Delta\theta_i^t$. The resulting dynamical equations for $i$th particle position $\mathbf{r}_i^t$ and velocity $\mathbf{v}_{i}^{t}$ are given by:}
\begin{eqnarray}
    \mathbf{r}_{i}^{t + \Delta t} &=& \mathbf{r}_{i}^{t} +\mathbf{v}_{i}^{t + \Delta t} \Delta t,
    \label{eq:updater_eqM}\\
    \theta_{i}^{t + \Delta t} &=& \theta_{i}^{t} + \Delta \theta^t_i + \xi_i^t,
    \label{eq:overdamped}
\end{eqnarray}
What remains is to choose a strategy to determine the reorientation angle $\Delta \theta^t_i$ in 
Eq.~\eqref{eq:overdamped}. In the classical discrete-time Vicsek model, $\Delta \theta^t_i$ is chosen to align the $i$th agent's velocity with the average velocity  $\mathbf{V}^t_i$ of its neighbors. To incorporate this effect, we define $\Delta \theta^t_i = \argmax_{\Delta \theta_i}  C_i^t$, i.e., as the argument that maximizes the correlation function
\begin{equation}
 C_{i}^t = \overline{\mathbf{v}}_i^{t+\Delta t}  \cdot \left( \sum_{j = 1}^{N} H\left(|\mathbf{r}_i - \mathbf{r}_j^t| -\zeta\right) \mathbf{v}_{j}^t - (\mathbf{v}_{i}^t -\mathbf{v}_{i})\right).
 \label{eq:eq5}
\end{equation}
\textcolor{black}{It can be interpreted as the correlation between the agent’s future desired velocity,
\(
\overline{\mathbf{v}}_{i}^{t + \Delta t} = v_0 \left[ \cos(\theta_i^t + \Delta \theta_i^t), \sin(\theta_i^t + \Delta \theta_i^t) \right],
\)
and the generalized, non-normalized average velocity of its predicted future neighbors within the interaction radius centered at its predicted future position $\mathbf{r}_i$ (see Fig.~\ref{fig:fig1}b). Since $C_{i}^t$ is not normalized, it quantifies the degree of alignment between the $i$th agent’s intended future heading and the prevailing orientation of its predicted future neighbors.}  Thus, it serves as a natural objective function to maximize by agents aiming to `copy' the prevalent orientation of their neighbors.  The Heaviside step function $H$ is modified such that $H(0) = 1$, ensuring that $C_{i}^t$ properly accounts for all particles within the interaction radius $\zeta$. \textcolor{black}{Depending on the cognitive abilities of the agents, the predicted velocity $\mathbf{v}_{i}$ and position $\mathbf{r}_{i}$ used in the non-normalized average velocity in Eq.~\eqref{eq:eq5} can be evaluated either at time $t$---for agents unable to predict their future state---or at time $t + \Delta t$---for cognitively more capable agents. This results in four distinct ways to define the correlation, as illustrated in Fig.~\ref{fig:fig1}b.
In principle, perceptual errors in real-world agents would necessitate the inclusion of a noise term within the bracket in Eq.~\eqref{eq:eq5}. However, we neglect such perceptual noise in the present study and, using the terminology of Vicsek model modifications, consider only angular noise while neglecting vectorial noise~\cite{Ginelli2016}.} 
If $C_i^t$ vanishes for all possible reorientations, the agent updates its orientation purely by noise, i.e., $\Delta\theta_i^t = 0$ in Eq.~\eqref{eq:overdamped}.

The strategies IA and IB calculate the correlation $C_{i}^t$ with the current neighbors of the agent $i$, $\mathbf{r}_i=\mathbf{r}_i^t$. Strategy IA further takes the agent's current velocity $\mathbf{v}_i=\mathbf{v}_i^t$ inside the sum, and IB uses the interpolated velocity $\mathbf{v}_i=\overline{\mathbf{v}}_i^{t+\Delta t}$ instead.  In both cases, $C_{i}^t = n_i^t \overline{\mathbf{v}}_i^{t+\Delta t}  \cdot \mathbf{V}^t_i + C_0$, where $C_0$ is a constant, $n^t_i$ the number of neighbors of agent $i$ at time $t$ and $\mathbf{V}^t_i$ their average velocity. For IA, $C_0=0$ and the agent $i$ is counted in $n^t_i$ and $\mathbf{V}^t_i$, so that $n^t_i = \sum_{j = 1}^{N} H\left(|\mathbf{r}_i^t - \mathbf{r}_j^t| - \zeta\right)$  and \textcolor{black}{$\mathbf{V}^t_i = \sum_{j = 1}^{N} H\left(|\mathbf{r}_i^t - \mathbf{r}_j^t| - \zeta\right) \mathbf{v}^t_j/n^t_i$. For IB, $C_0=v_0$} and the agent $i$ is not counted in the definition of $n^t_i$ and $\mathbf{V}^t_i$ ($j\neq i$ in the sums above). Nevertheless, in both cases, $n_i^t$ and $C_0$ are independent of 
$\Delta \theta_i^t$ and thus the intended velocity that maximizes $C_{i}^t$ is the one best aligned with 
the average velocity $\mathbf{V}^t_i$.
Notably, considering the agent's own velocity in \( \mathbf{V}_i^t \) introduces slight orientational inertia in IA, as agents take their own heading into account.
 These two strategies correspond to two variants of the Vicsek model: Vicsek model A, which includes the agent's own velocity in the average velocity calculation, and Vicsek model B, which does not (see Sec.~S1~\cite{supplementary}).

The strategies IIA and IIB, use the neighbors corresponding to the intended future position of agent $i$ at time $t+\Delta t$, $\mathbf{r}_i=\mathbf{r}_i^t  + \overline{\mathbf{v}}_i^{t+\Delta t} \Delta t$, and thus require calculating the correlation $C_{i}^t$ using different neighbors for each value of the realignment angle. From now on, we will call these two strategies predictive and the corresponding models as predictive models. As above, strategy IIA further takes the agent's current velocity  $\mathbf{v}_i=\mathbf{v}_i^t$ inside the sum, and IIB the interpolated velocity $\mathbf{v}_i=\overline{\mathbf{v}}_i^{t+\Delta t}$. Also in these cases, $C_{i}^t = n_i^t \overline{\mathbf{v}}_i^{t+\Delta t}  \cdot \mathbf{V}^t_i + C_0$. Nevertheless, the number of neighbors of $i$ , $n_i^t$, and their average velocity, $\mathbf{V}^t_i$, are now calculated with respect to its intended position $\mathbf{r}_i(t)+\overline{\mathbf{v}}_i(t+\Delta t)\Delta t$ and thus they depend on the reorientation angle. For IIA the agent $i$ is counted in $n_i^t$ and $C_0 = 0$. \textcolor{black}{For IIB, $C_0 = v_0$} and the agent $i$ does not contribute to the averages. Importantly, in both these strategies, the optimal reorientation angle follows from a tradeoff balancing the number of nearest neighbors and alignment with the average velocity, resulting in an attractive alignment interaction. Different from IIB, IIA, in addition, has some positional inertia.

The time step \( \Delta t \) affects only the relaxation times and does not alter the stationary state. Upon rescaling particle positions by the interaction radius $\zeta$, the stationary behavior of this model is controlled by two parameters: the ratio of the distance traveled per timestep to the interaction radius, \( v_0 \Delta t / \zeta \), and the noise-induced orientation change per time step, quantified by \( \eta \). In the following, we consider groups of \( N = 200 \) agents initially positioned randomly within a square of side length \( L = 4\zeta \), with \( \zeta = 1 \) and \( \Delta t = 1 \). In Sec.~S10~\cite{supplementary}, we show that using a larger $N = 500$ produces qualitatively the same results. A more physically grounded, continuous-time variant of the model is described in Sec.~S2~\cite{supplementary}.

\emph{Flocking from predictive alignment}:
\begin{figure*}[ht!]
\centering
\begin{minipage}[b]{7in}
    \includegraphics[width=0.7
    \textheight,center]{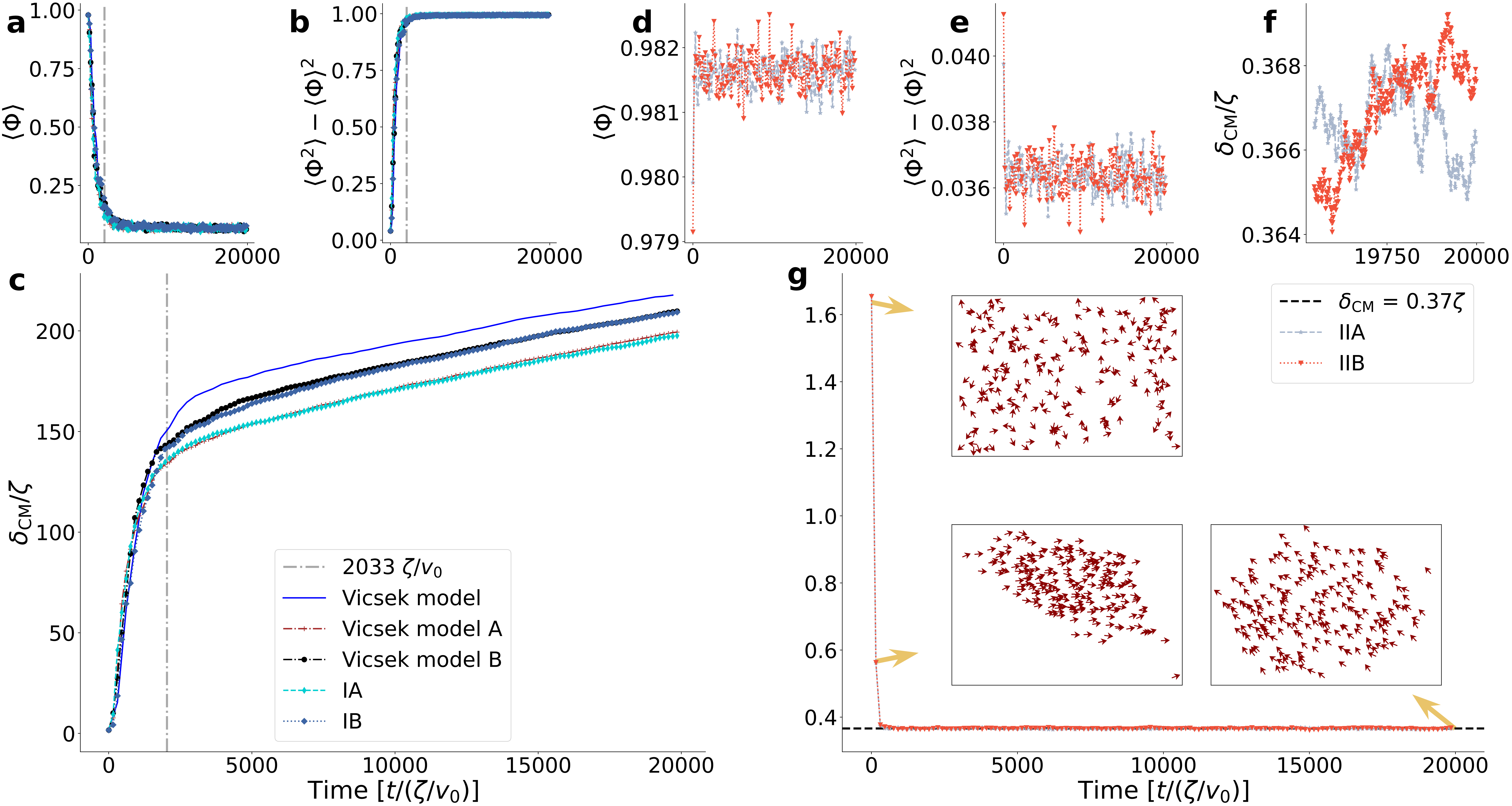}
\caption{\textbf{Comparison between Vicsek-like and predictive models.} 
The agents started with uniformly distributed orientations and evolved according to the standard time-discrete Vicsek model, its modifications A and B (see Sec.~S1~\cite{supplementary}), as well as the decision-based models IA–IIB defined in the main text.
For all models, we set the reduced speed to $v_0/\zeta = 0.0076$, noise intensity to $\eta = 0.1$, and averaged the shown data over 25 replicas with different noise realizations. \textbf{a}–\textbf{c}, The Vicsek-type models exhibit a rapid loss of cohesion, indicated by \textbf{a} a sharp decrease in the average polarization $\langle \Phi \rangle$, \textbf{b} an increase in its fluctuation $\langle \Phi^2 \rangle - \langle \Phi \rangle^2$, and \textbf{c} a rapid growth of the average agent-to-center-of-mass distance, $\delta_{\rm CM}$. These effects occur before $t \approx 2033 \zeta / v_0$, predicted from diffusive spreading analysis of Vicsek model (vertical dashed lines). \textbf{d}–\textbf{g}, The predictive models IIA and IIB yield nearly identical stable flocking behaviors, with \textbf{d} a consistently high average polarization fluctuating weakly around 0.98, \textbf{e} low polarization variance, and \textbf{f} a closely packed system configuration, where the average agent-to-center-of-mass distance fluctuates around 0.366. \textbf{g} The system size self-adjusts as the initially square-shaped flock transitions through an elongated intermediate state before settling into a final circular configuration (insets). The system size relaxation time, defined as the point when $\delta_{\rm CM}$ drops to half of its initial value, is approximately $100\zeta/v_0$. Analogously defined relaxation times for $\langle\Phi\rangle$ and $\langle\Phi^2\rangle - \langle\Phi\rangle^2$ are shorter than $\zeta/v_0$ (see Sec.~S7~\cite{supplementary}).}
\label{fig:fig2}
\end{minipage}
\end{figure*}
Models with purely alignment interactions, such as the Vicsek model, fail at maintaining group cohesion even under arbitrarily weak noise due to the diffusive spreading of agents.  The time it takes for two particles, initially at the same position, to `diffuse' further away than one interaction radius can be estimated as $\left(\frac{1}{5} + \frac{3}{2 \pi^2 \eta^2}\right) \frac{\zeta^2}{v_0^2}$ (see Sec.~S4~\cite{supplementary}).
It is reasonable to expect that as the number of agents increases, the Vicsek flock will break up into subgroups more quickly. For $\eta = 0.1$ and $v_0/\zeta = 0.0076$ as used in Fig.~\ref{fig:fig2}, our estimate suggests that flock coherence is lost before $t \approx 2033 \zeta / v_0$. For Vicsek-like models IA, IB, and for the standard Vicsek model, this prediction aligns remarkably well with the saturation point where (a) the average polarization, $\langle\Phi\rangle$, halts its rapid decrease, and (b) the polarization variance, $\langle\Phi^2\rangle - \langle\Phi\rangle^2$, halts its rapid increase. It also marks the end of the initial sharp rise in the average agent-to-center-of-mass distance, $\delta_{\rm CM}$ (c). Beyond this point, the system size expands ballistically as the single flock fragments into multiple sub-flocks, indicated by the vanishing polarization and peak variance in (a) and (b). (For precise definitions of the order parameters, see Sec.~S3~\cite{supplementary}.)
\textcolor{black}{On the other hand, the predictive models IIA and IIB produce highly polarized, closely packed, and coherent flocks, with a self-adjusted $\delta_{\rm CM} \approx 0.4 \zeta$, corresponding to a flock radius of approximately $0.6 \zeta$. This implies that the entire stationary flock fits within a single interaction radius, making the model unrealistic from a biological perspective, where the number of perceived neighbors is limited~\cite{Ballerini2008}. The corresponding order parameters exhibit only minor fluctuations and remain stable over time—at least over the simulation durations we tested, which span up to ten diffusive spreading times.
} 

\emph{Noise induced dynamical transition}:
\begin{figure*}[ht!]
\centering
\begin{minipage}[b]{7in}
    \includegraphics[width=0.9
\textwidth,center]{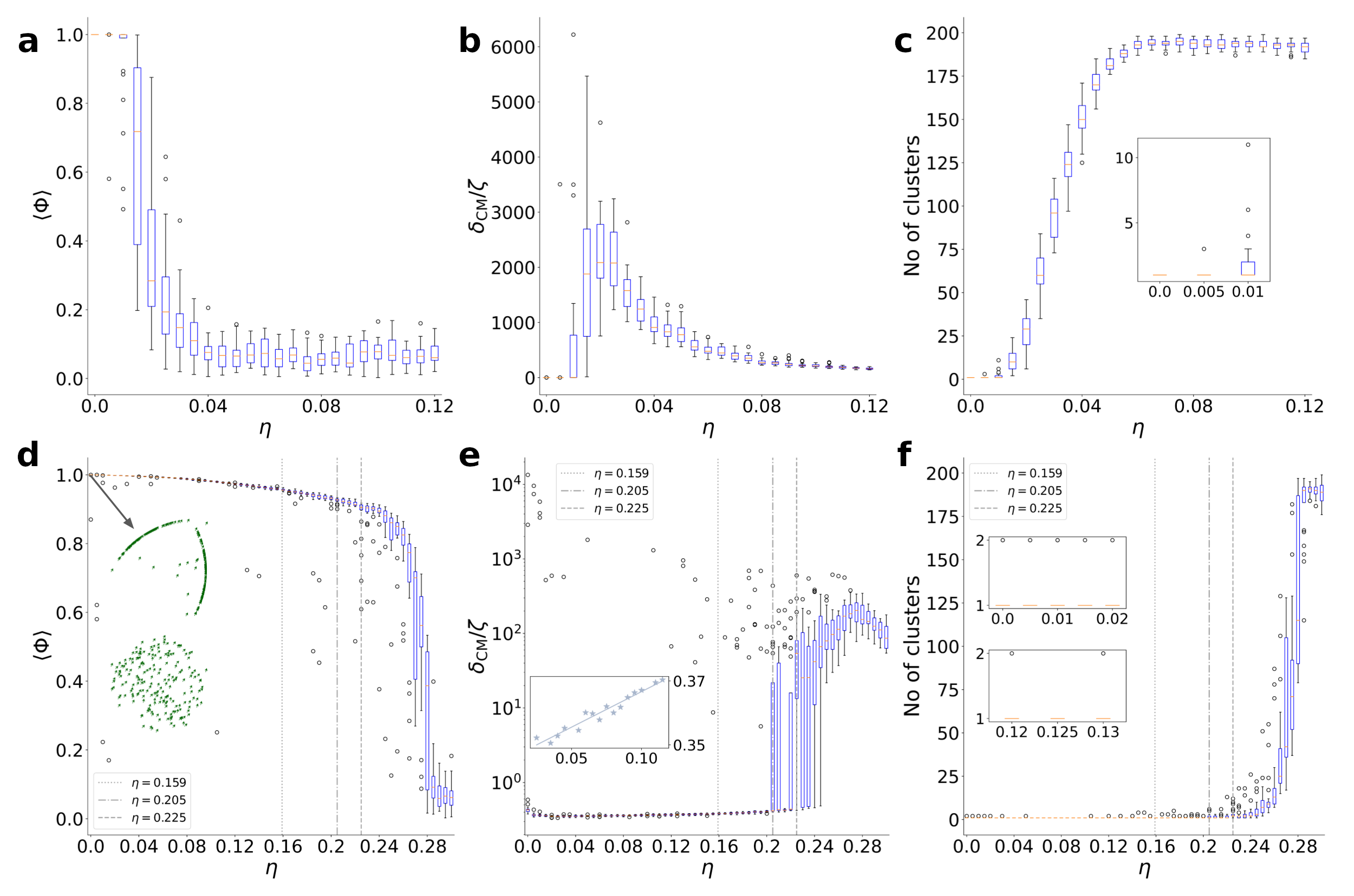}
\caption{\textbf{Effects of noise in Vicsek model and the predictive model IIA}. Boxplots represent results from 25 independent simulations with different noise realizations, where orange lines indicate the median, boxes span the interquartile range, whiskers extend to data points within 1.5 times the interquartile range, and outliers are shown as individual circles. \textbf{a}–\textbf{c} In the Vicsek model, for noise levels $\eta \gtrapprox 0.005$, flock cohesion is lost, with fragmentation increasing at higher noise levels, as reflected by \textbf{a} reduced average polarization $\langle \Phi \rangle$, \textbf{b} an increased average agent-to-center-of-mass distance $\delta_{\rm CM}$, and \textbf{c} a greater number of clusters. \textbf{d}–\textbf{f} In contrast, the predictive model IIA maintains stable flocking in over half of the replicas for $\eta \lessapprox 0.225$. Here, \textbf{d} polarization gradually decreases from 1, with agents forming V-shaped flocks at zero noise and round flocks at nonzero noise (insets). \textbf{e} The average agent-to-center-of-mass distance initially decreases but subsequently increases linearly for $0.015 \lessapprox \eta \lessapprox 0.225$ and coherent replicas, following $\delta_{\rm CM} \approx (0.35 + 0.19\eta)\zeta$ (inset). \textbf{f} The system predominantly consists of a single cluster of communicating agents for $\eta \lessapprox 0.225$, with more than one outlier for $0.015 \lessapprox \eta$ and $0.159 \lessapprox \eta \lessapprox 0.205$. At high noise levels, system size (\textbf{b}, \textbf{e}) decreases due to the interplay between noise-induced alignment destabilization and suppression of system growth by the diffusive motion of individual subclusters. The models were simulated under the same conditions as in Fig.~\ref{fig:fig2} unless otherwise specified in the figure. The order parameters were evaluated at time $2\times 10^4 \zeta/v_{0}$.}
    \label{fig:fig3}
\end{minipage}
\end{figure*}
The predictive strategies IIA and IIB yield nearly identical results, while strategy IA exhibits slightly better coherence than IB. We attribute this to the slight orientational inertia introduced by the definition of the correlation function in strategies A. In the following, we analyze the behavior of the IIA model under variations in the two key parameters: noise intensity, $\eta$, and scaled velocity, $v_0/\zeta$.

With periodic boundary conditions~\cite{Vicsek1995}, the Vicsek model undergoes a discontinuous phase transition~\cite{ChateH-2008} from an ordered to a disordered state. Without periodic boundaries, coherent polarized flocks form only at vanishing noise. When initialized with randomly oriented agents uniformly distributed within a rectangle of side length $4\zeta$, the model exhibits a monotonic decrease in average polarization (Fig.~\ref{fig:fig3}a) and a corresponding increase in the number of communicating clusters (Fig.~\ref{fig:fig3}c) as noise intensifies, consistent with this expectation. Notably, the average agent-to-center-of-mass distance reaches a maximum at an intermediate noise level (Fig.~\ref{fig:fig3}b). This nonmonotonic behavior arises because, at low noise, the flock expands ballistically, whereas at high noise, the motion of individual subflocks becomes diffusive on the relevant timescale. In this regime, subflocks undergo an effective random walk, slowing the overall expansion of the system.  

Under the same conditions and for noise intensities $\eta \lessapprox 0.225$, the predictive model IIA produces coherent flocks consisting of a single cluster of communicating particles (Fig.\ref{fig:fig3}f) with polarization $\langle \Phi \rangle \approx 1$ (Fig.\ref{fig:fig3}d) in most of the 25 replicas used in our simulations. The inset shows that, in the absence of noise, the coherent flocks adopt a V-shaped formation, reminiscent of those observed in migrating birds, where this arrangement reduces energy expenditure. At nonzero noise levels, the flocks transition to a rounded shape, similar to the formations observed in foraging bird flocks, where cohesion and flexibility are prioritized over aerodynamic efficiency. For videos showing the relaxation of flock shapes and an analysis of the corresponding relaxation times, see the SI~\cite{supplementary}.

For $\eta \lessapprox 0.015$, the average agent-to-center-of-mass distance decreases with increasing noise. This `noise stabilization effect' arises from the discrete set of allowed reorientations, which, unlike the classical Vicsek model with arbitrary reorientation per timestep, prevents the system from fully polarizing at zero noise. A similar effect has been observed in Ref.~\cite{Turner2023}. For $0.015 \lessapprox \eta \lessapprox 0.225$, the average agent-to-center-of-mass distance in stable replicas increases linearly with noise (inset of Fig.~\ref{fig:fig3}e). Beyond $\eta \approx 0.225$, all order parameters undergo a transition for the majority of replicas: polarization $\langle \Phi \rangle$ vanishes, $\delta_{\rm CM}$ grows by two orders of magnitude within the given simulation time, and the number of clusters approaches the total number of agents. At higher noise levels, both $\delta_{\rm CM}$ and the number of clusters slightly decrease, consistent with the diffusive motion of subclusters described above.

In the 25 replicas of the system with different noise realizations obtained from our simulations, a few exceptions to the described behavior appear as empty circles in Fig.~\ref{fig:fig3}, representing individual outliers from the typical trend, depicted by the orange lines inside the boxes. The higher number of outliers observed for $\eta \approx 0$ in Fig.~\ref{fig:fig3}e, compared to Fig.~\ref{fig:fig3}f, arises because each replica contributing to the system size outliers consisted of two separate subflocks, leading to overlapping circles in Fig.~\ref{fig:fig3}f. In Sec.~S9~\cite{supplementary}, we show that the same phenomenology can also be observed when the system is initially perfectly aligned, demonstrating the robustness of the described dynamic phases. For further details on how the described dynamical phases manifest in the behavior of the individual replicas, see Fig.~S1~\cite{supplementary}. 

\emph{Role of speed and interaction radius}:
For a given noise intensity, the system forms a stable flock if the ratio \( v_0/\zeta \) is small enough so that each agent has sufficient time to align with its neighbors before changing them. In the stable regime, the flock size is proportional to the interaction radius and independent of the speed, i.e., \( \delta_{\rm CM} \sim \zeta_s \). For details, see Fig.~S1~\cite{supplementary}.

\emph{Leadership}:
In nature, bird flocks often involve a subgroup of leaders who are best informed about the target position and who are followed by the rest of the flock~\cite{Couzin2005,Nagy2010}. In Fig.~S9 and Supplementary video 3~\cite{supplementary}, we show that the predictive model IIA can form cohesive flocks also in the scenario when a subgroup of leaders perform an oscillator deterministic motion, albeit for slightly lower $v_0/\zeta$ than without the perturbation by leaders.

\emph{Discussion}:
\textcolor{black}{We have presented a cohesive flocking model based solely on alignment interactions, achieved by replacing the Ising-like alignment rule of the Vicsek model with \emph{predictive alignment}, in which agents adopt the predicted prevailing orientation of their future neighbors. For agents unable to predict their future positions, this rule reduces to various modifications of the Vicsek model—since the set of neighbors remains the same for all directions, the magnitude of the mean polarization is independent of the chosen direction. However, agents that can predict the future positions of their neighbors optimize a tradeoff between aligning with neighbors’ headings and maintaining proximity, yielding cohesion and order without the need for boundaries or added forces. This approach fundamentally departs from previous models, which rely on such aids~\cite{Chate2008b}, and is reminiscent of the reinforcement learning algorithm aimed at minimizing neighbors’ losses, as investigated in Refs.~\cite{Durve2020, brambati2025learningflockopenspace}.} From a technical perspective, the dynamical equations feature a reorientation 'force' that does not follow the gradient of a potential, which would typically lead to stable orientations at local minima. Instead, it is governed by an 
$\argmax$ function, which reorients agents toward the deepest minimum of a utility function (negative orientation correlation with neighbors) that is accessible in the next timestep. This process is constrained by the agent’s field of view, reorientation capabilities, and motility. 

\textcolor{black}{
Our algorithm provides a plausible strategy that intelligent agents with given physical and cognitive abilities might employ to efficiently align with their neighbors. As such, it falls within the class of intrinsically motivated \cite{MehdiCrowds2011, RomanczukPursuitEscape2009, Charlesworth2019} and cognitive \cite{Barberis2016, LavergneBechingerScience2019} active matter algorithms. The algorithm can also be integrated into the broader framework of active inference \cite{ActiveInference2022}, a general theory of decision-making. However, unlike typical active inference models, our approach does not rely on the assumption that the system state is near the global optimum of a utility function, allowing forces to be described as gradients of generalized potentials. Instead, it enables agents to dynamically adapt the most preferred configuration they perceive.}

The model is scalable, and the resulting flock shapes resemble those observed in nature.
\textcolor{black}{However, the stationary states predicted by the model are so dense that the average number of neighbors perceived by each agent is significantly higher than the realistic values natural agents are able to process—typically around seven~\cite{Ballerini2008}. Moreover, birds have been shown to align with their nearest topological, rather than metric, neighbors~\cite{Ballerini2008}. In addition to these issues, future revisions of the model should be accompanied by an analysis of the properties commonly studied in natural flocks or swarms, such as the shapes of correlation functions and their finite-size scaling~\cite{Cavagna2018}, to allow for a quantitative comparison between natural systems and the model.
}

\textcolor{black}{Future extensions of the model could investigate modifications to agents’ cognitive abilities—such as enhanced predictive capabilities, perceptual limitations~\cite{Ballerini2008}, or delays in decision-making processes~\cite{Holubec2021}. Another avenue is to consider agents governed by different physical principles, for example, incorporating inertia or more general non-Markovian effects. Finally, it would be valuable to explore potential applications of models like the one presented here in areas such as swarm robotics~\cite{hamann2018swarm}, where agents are not constrained by biological limitations. Notably, the current approach resembles swarm control algorithms based on individual robot decisions made without explicit information sharing among agents~\cite{BAYINDIR2016292}.}


\section*{Acknowledgments}

This work is supported and financed by Charles University in Prague, project PRIMUS/22/SCI/009. We acknowledge Klaus Kroy and Maurice Zeuner for their contributions to the initial version of the project, which is summarized in Maurice Zeuner’s bachelor's thesis, defended at the University of Leipzig in 2020.

\bibliography{Main}

\end{document}


\flushbottom
\maketitle
\thispagestyle{empty}
\renewcommand{\figurename}{Supplementary Figure}

\section{Variants of the Vicsek model}
\label{ssec:mod_vm}

In the original discrete-time variant of the Vicsek model~\cite{Vicsek1995}, agent positions are updated according to Eq.~(1) in the main text, while their velocities $\mathbf{v}_{i}^{t+1}$ are determined by the average velocity of their neighbors at time $t$,  
$\mathbf{V}_i = \sum_{j = 1}^{N} H_x\left(|\mathbf{r}_i^t - \mathbf{r}_j^t| -\zeta\right) \mathbf{v}_{j}^t$,  
where $\mathbf{V}_i$ is then randomly rotated by an angle $\xi_i^t$, as described in Eq.~(2) in the main text. 
Here, $H_x$ (with $x = A, B$) are Heaviside theta functions modified at the origin such that agent $i$'s own velocity is included ($H_A(0) = 1$) or excluded ($H_B(0) = 0$) in the averaging, yielding variants of the model with slight orientational "inertia" or no inertia, respectively.

We compare the `predictive' models defined in the main text with the original Vicsek model using $H_A$. However, this comparison is not entirely fair, as the predictive models do not allow for arbitrary reorientation within a single time step. To ensure a fair comparison, we also compare the predictive models with variants of the Vicsek model using $H_A$ or $H_B$, where agents reorient by the angle in the set $\Omega_\theta$ that makes their velocity closest to $\mathbf{V}_i$ before undergoing random reorientation due to noise. We call these variants of the Vicsek model as Vicsek model A and B, respectively. They are identical to the models corresponding to strategies IA and IB defined in the main text.

\section{Time-continuous model}
\label{ssec:continuous_model}

Physically, the algorithmic discrete-time model in Eqs.~(1) and (2) in the main text is reasonable when agents travel only a fraction of the interaction radius per time step, i.e., $v_0 \Delta t \ll \zeta$, ensuring that they do not switch neighbors at each step. This condition is fulfilled in all our numerical experiments.  
In this parameter regime, one can readily take the continuous-time limit $\Delta t \to 0$ in Eq.~(1) to obtain  
\begin{equation}
\dot{\mathbf{x}}_i(t) = \mathbf{v}_i(t).
\label{eq:cont}
\end{equation}  
Introducing a reorientation angular velocity, $\omega_0$, and rotational diffusion, $D_r$, the continuous-time variant of Eq.~(2) can be formulated as  
\begin{equation}
\dot{\theta}_{i}(t) =  \omega_0 \Delta \theta_i(t) + \sqrt{2 D_r} \xi_i(t),
\end{equation}  
where $\xi_i(t)$, $i=1,\dots,N$, are normalized, unbiased, and mutually independent Gaussian white noises. \textcolor{black}{We stress that this is not the time-continuous limit of the time-discrete model considered in the main text. Rather, it is a reasonable time-continuous variant formulated using the same logic as the time-continuous versions of the Vicsek model.}
 
\section{Order parameters}
\label{ssec:order_parameters}

We characterize the studied systems using the average polarization $\langle\Phi\rangle$, polarization variance $\langle \Phi^{2} \rangle - \langle \Phi \rangle^{2}$, the average agent-to-center-of-mass distance~\cite{Kattas2012}, $\delta_{\rm CM}$, which serves as a proxy for system size, and the number of clusters of communicating particles. These variables are calculated as 
\begin{eqnarray}
\langle \Phi \rangle &=& \frac{1}{v_0} |\langle \mathbf{v} \rangle| =  \frac{1}{v_0 N}\bigg| \sum_{i=1}^{N} \mathbf{v}_i \bigg|, \\
\langle \Phi^{2} \rangle - \langle \Phi \rangle^{2} 
&=& \frac{1}{v_0^2 N}\sum_{i=1}^{N}\left( \mathbf{v}_i-\langle \mathbf{v} \rangle \right)^2, \\
\delta_{\rm CM} &=& \sqrt{\frac{1}{N} \sum_{i=1}^{N} \left| \mathbf{x}_{i}(t) - \left\langle \mathbf{x}(t) \right\rangle \right|^{2}},
\end{eqnarray}
with the flock center of mass position vector $\left\langle \mathbf{x}(t) \right\rangle = \frac{1}{N} \sum_{i=1}^{N} \mathbf{x}_{i}(t)$.
 The number of clusters is calculated by iteratively identifying all particles that can be connected through a path where each step links particles separated by a distance smaller than the interaction radius $\zeta$.

For a homogeneous circular flock with radius $R$, 
$\delta_{\rm CM} = \frac{1}{\pi R^2} \int_0^{2\pi} d\phi \int_0^R r \, dr \, r = \frac{2}{3}R$. 
This result can be used to estimate the flock radius from the easily calculable $\delta_{\rm CM}$.

\section{Diffusive spreading in Vicsek model}
\label{ssec:diffusive_spreading}

\textcolor{black}{To estimate the speed of the inevitable noise-induced spreading of agents in the Vicsek model, we consider two particles interacting via a perfect, infinite-range alignment interaction. At each time step, they align their velocities and add a uniformly distributed noise term $\xi_i \in \eta[-\pi, \pi]$ to their orientation. Consequently, after each time step, the velocities of the two particles are given by $\mathbf{v}_i = v_0(\cos \theta_i, \sin \theta_i)$ for $i = 1,2$, with the angular difference given by $\theta_1 - \theta_2 = \xi_1 - \xi_2$.
 Per time step, the distance between the two particles increases by $\Delta d = |\mathbf{v}_1 - \mathbf{v}_2|\Delta t = v_0 \Delta t \left|\left(\cos\xi_1-\cos\xi_2, \sin\xi_1-\sin\xi_2\right)\right|$, $\Delta t = 1$. Since these individual distance increments are independent by construction, the probability density of the distance between the two particles after a large number of time steps $t$ can be well approximated by a Gaussian distribution with zero mean and variance  
\[
\langle\Delta d^2\rangle t =
\frac{v_0^2 t}{(2 \pi \eta)^2} \int_{-\eta\pi}^{\eta \pi}d\xi_1 \int_{-\eta\pi}^{\eta \pi}d\xi_2\,\left[2 - 2 \cos(\xi_1 - \xi_2)\right] =
\frac{v_0^2 t \left(2 \pi^2 \eta^2 - 1 + \cos(2 \pi \eta) \right)}{\pi^2 \eta^2},
\]
where the average is taken over the noises $\xi_1$ and $\xi_2$.} This result provides an estimate for the expected distance between the two particles after $t$ time steps as $\sqrt{\langle\Delta d^2\rangle t}$. Similarly, the time at which the distance between the two particles exceeds the interaction radius $\zeta$ — marking the loss of coherence in the system — can be estimated as 
$t \approx \frac{\zeta^2}{\langle\Delta d^2\rangle}$.

\textcolor{black}{We note that conceptually similar estimates can also be made for the spreading of angular perturbations in the ordered phase of the Vicsek model~\cite{toner2024physics}. }

\section{Replica-resolved behavior of the predictive model IIA with respect to noise}
To provide further insight into the behavior of individual replicas, Fig.~\ref{fig:fig4}a shows the logarithm of the average agent-to-center-of-mass distance. Dark red-colored replicas indicate a small stationary system size and thus stability, whereas blue and faint red mark unstable replicas. The reduced number of blue points beyond $\eta \le 0.015$ illustrates the aforementioned noise-induced stabilization effect. The figure also shows that for $0.159 \lessapprox \eta \lessapprox 0.205$, the number of unstable replicas sharply increases. For $0.205 \lessapprox \eta \lessapprox 0.225$, unstable replicas are no longer mere outliers, and for $\eta \gtrapprox 0.225$, the system becomes unstable in the majority of replicas. Interestingly, the very onset of the transition at $\eta \approx 0.159$ corresponds to the parameter regime when the maximum 
‘intentional’ reorientation of the agents, $0.5$ rad, matches the 
maximum reorientation due to noise, $\eta\pi$. 

\section{Role of speed and interaction radius:}
\begin{figure*}[ht!]
\centering
\begin{minipage}[b]{7in}
    \includegraphics[width=0.8
\textwidth,center]{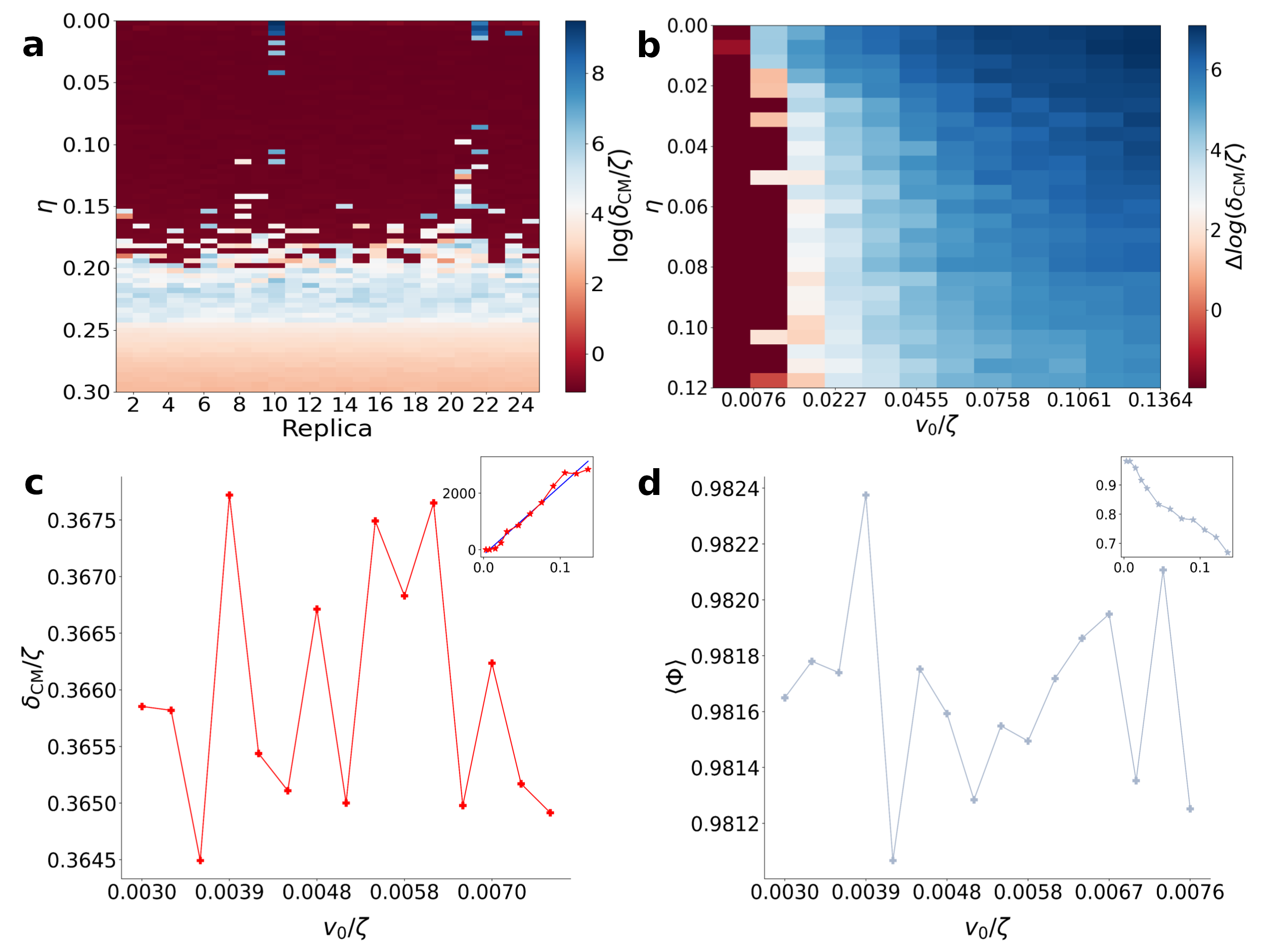}
    \caption{\textbf{Effects of noise, speed, and interaction radius in the predictive model IIA}. 
    \textbf{(a)} The logarithm of the average agent-to-center-of-mass distance, $\log (\delta_{\rm CM}/\zeta)$, for individual replicas as a function of noise intensity, $\eta$.  
\textbf{(b)} Change in $\log (\delta_{\rm CM}/\zeta)$ between the final simulation time, $2 \times 10^4 \zeta/v_0$, and an earlier time, $2 \times 10^4 (\zeta/v_0 - 1)$, as a function of $\eta$ and the fraction of the interaction radius traveled per time step, $v_0/\zeta$.  
\textbf{(c)} The average agent-to-center-of-mass distance as a function of the reduced speed, $v_0/\zeta$, in the regime where the model forms stable flocks (taking out outliers).  
The insets in \textbf{(c)} and \textbf{(d)} show $\delta_{\rm CM}$ and the average polarization, $\langle\Phi\rangle$, as functions of $v_0/\zeta$ outside the stable regime.  
The models were simulated under the same conditions as in Fig.~2 in the main text, with $\eta = 0.1$ and $v_0/\zeta = 0.0076$, unless otherwise specified in the figure. The order parameters were evaluated at time $2\times 10^4 \zeta/v_0$.
} 
    \label{fig:fig4}
\end{minipage}
\end{figure*}

In Fig.~\ref{fig:fig4}b-d, we analyze the distance traveled per timestep over the interaction radius, $v_0/\zeta$, on the system dynamics. The system forms a stable flock if the fraction is small enough so that each agent has enough time to align with its neighbors before it changes them. The threshold value of $v_0/\zeta$ increases with noise intensity from roughly 0.008 at $\eta \lessapprox 0.02$ to almost 0.024 at $\eta = 0.12$ (Fig.~\ref{fig:fig4}b), highlighting once more the stabilizing effect of the noise discussed above. In the stable regime, the average agent-to-center-of-mass distance reduced by $\zeta$ (Fig.~\ref{fig:fig4}c) and the average polarization (Fig.~\ref{fig:fig4}d) are independent of $v_0/\zeta$. Hence, the system size is proportional to the interaction radius. Beyond the stable regime, $\langle\Phi \rangle$ drops and $\delta_{\rm CM}$ increases with both $v_0/\zeta$ and the simulation time.

\section{Relaxation of flock shapes in predictive model IIA}

\begin{figure*}[ht!]
\begin{minipage}[b]{7in}
\centering
    \includegraphics[width=0.65
\textwidth,center]{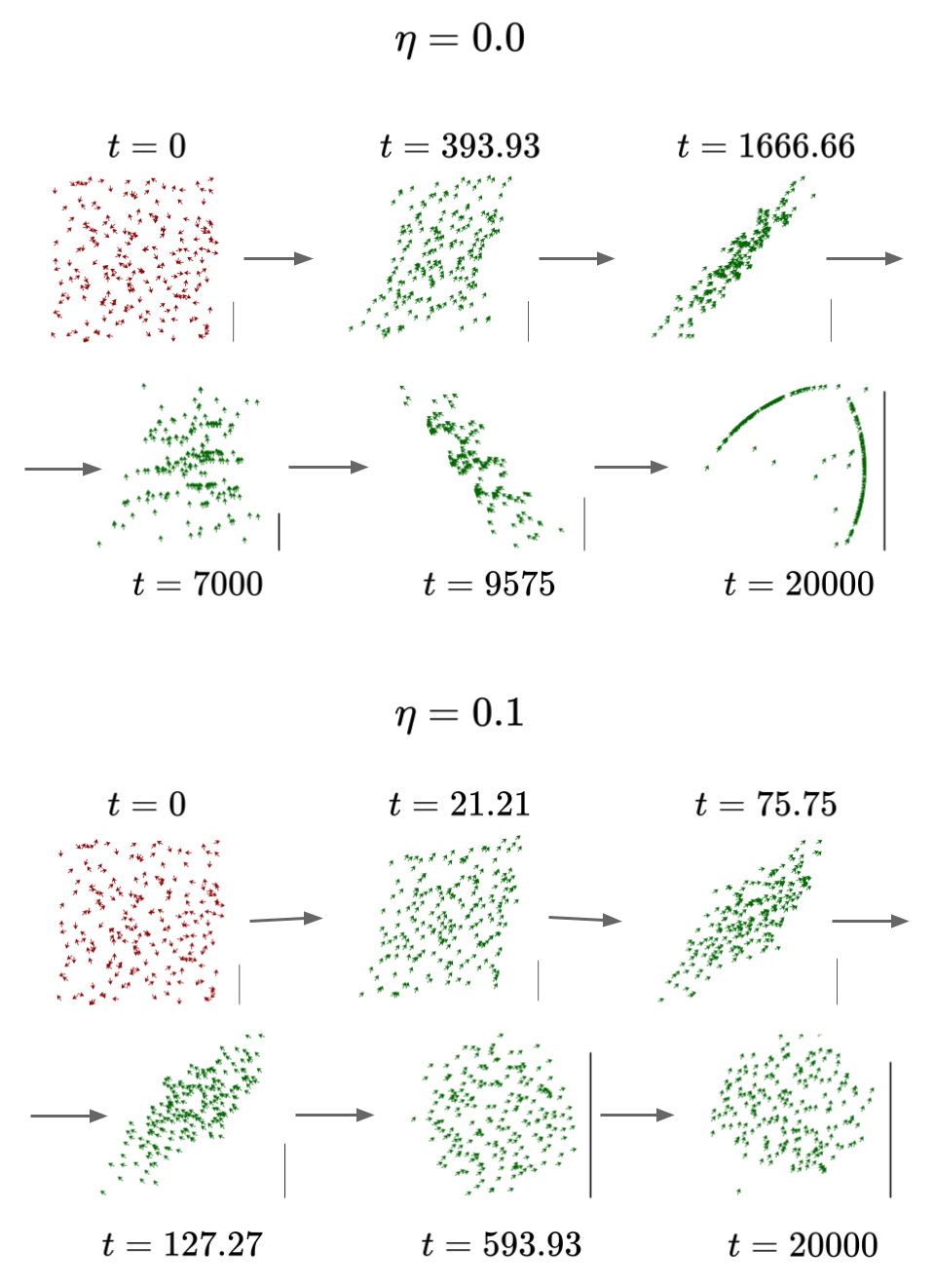}
\caption{\textbf{Relaxation dynamics of the predictive model IIA} 
The top two rows of the figures show the relaxation of the system for vanishing noise, where the final stationary state is a V-shaped flock. The bottom two rows show the same for a noise intensity of $\eta = 0.1$ when the stationary state is a circular flock. The vertical scale bars at the bottom-right of all panels have a length of one interaction radius, $\zeta = 1$, and illustrate how the system shrinks over time. In both cases, we simulated 200 agents initialized with uniformly distributed orientations in $[0, 2\pi]$ and positions within a box of size $4\zeta \times 4\zeta$. The agents' reduced speed is $v_0/\zeta = 0.0076$.}
\label{fig:SI_fig1}
\end{minipage}
\end{figure*}

As shown in the main text, for sufficiently low noise and a small enough ratio of agent speed to interaction radius, the predictive model IIA converges to a steady state. In this steady state, the system forms V-shaped flocks for vanishing noise and round flocks for nonzero noise (see Fig.~3d in the main text). In Fig.~\ref{fig:SI_fig1}, we additionally illustrate how the system relaxes from its initial square shape with randomly oriented agents to the two stationary shapes using noise intensities $\eta = 0.0$ and $0.1$.

\begin{figure*}
\centering
\begin{minipage}[b]{7in}
\includegraphics[width=0.9 \textwidth,center]{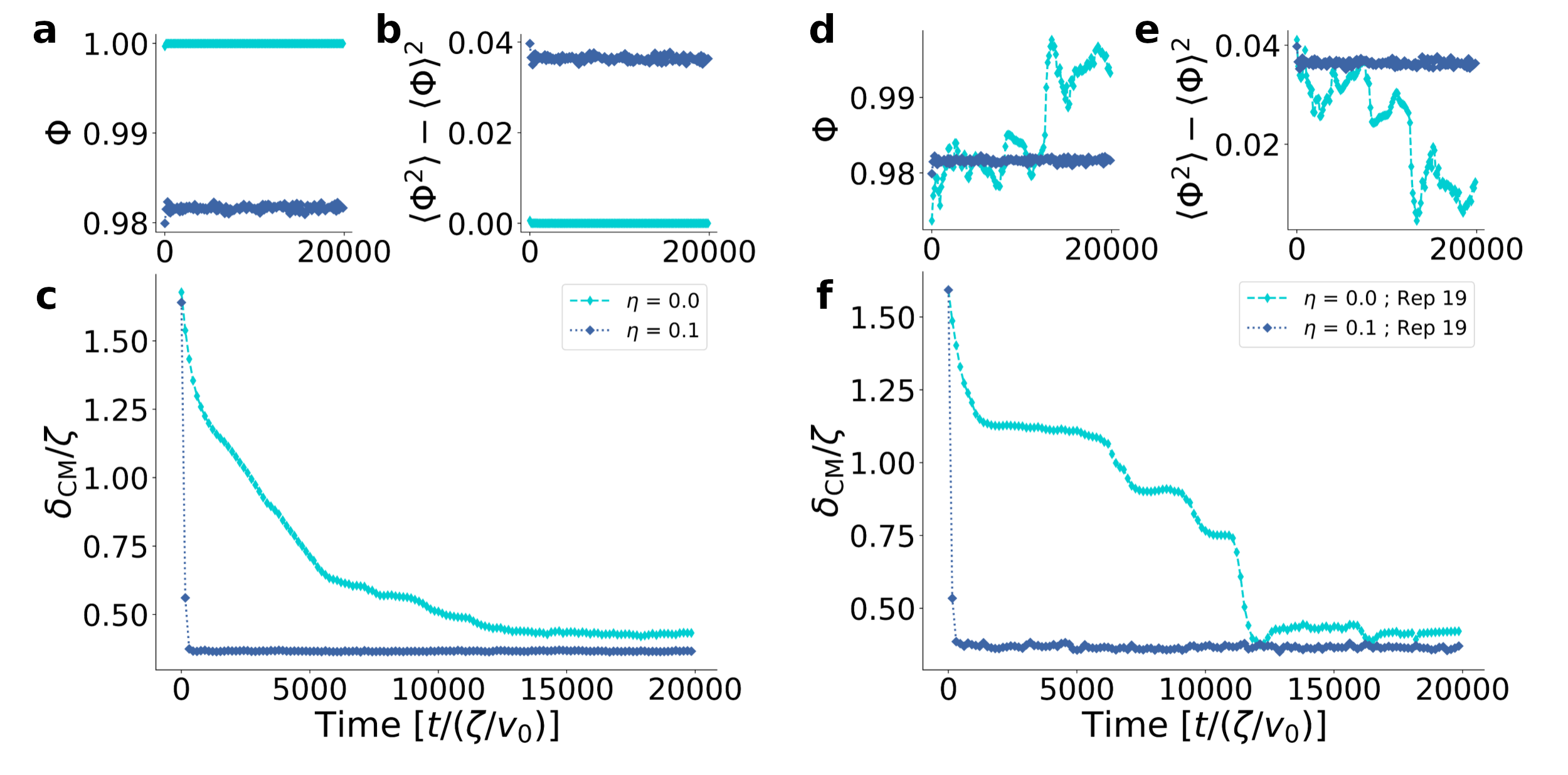}
\caption{\textbf{Relaxation of order parameters for the predictive model IIA.} (\textbf{a}-\textbf{c}) The time evolution of \textbf{a} average polarization, $\langle\Phi\rangle$, \textbf{b} polarization variance, $\langle\Phi^2\rangle-\langle\Phi\rangle^2$, and \textbf{c} average agent-to-center-of-mass distance, $\delta_{\rm CM}$ for the same parameters as used in Fig.~\ref{fig:SI_fig1}, averaged over stable replicas from 25 simulated replicas. (\textbf{d}-\textbf{f}) The same for the replicas used to produce Fig.~\ref{fig:SI_fig1}.
}
\label{fig:SI_fig2}
\end{minipage}
\end{figure*}

To gain further insight into the two relaxation processes, we also show in Fig.~\ref{fig:SI_fig2} the corresponding relaxation of average polarization, $\langle\Phi\rangle$, polarization variance, $\langle\Phi^2\rangle-\langle\Phi\rangle^2$, and average agent-to-center-of-mass distance, $\delta_{\rm CM}$. Specifically, Fig.~\ref{fig:SI_fig2}a-c show the relaxation of these quantities for $\eta = 0.0$ and $0.1$, averaged over all stable replicas. Figures~\ref{fig:SI_fig2}d-f then show the relaxation for the two replicas used to produce snapshots in Fig.~\ref{fig:SI_fig1}. While the relaxation trajectories for the noisy system are smooth, those corresponding to vanishing noise exhibit `metastable plateaus' interconnected by fast relaxations.

\begin{figure*}
\centering
\begin{minipage}[b]{7in}
\includegraphics[width=0.85 \textwidth,center]{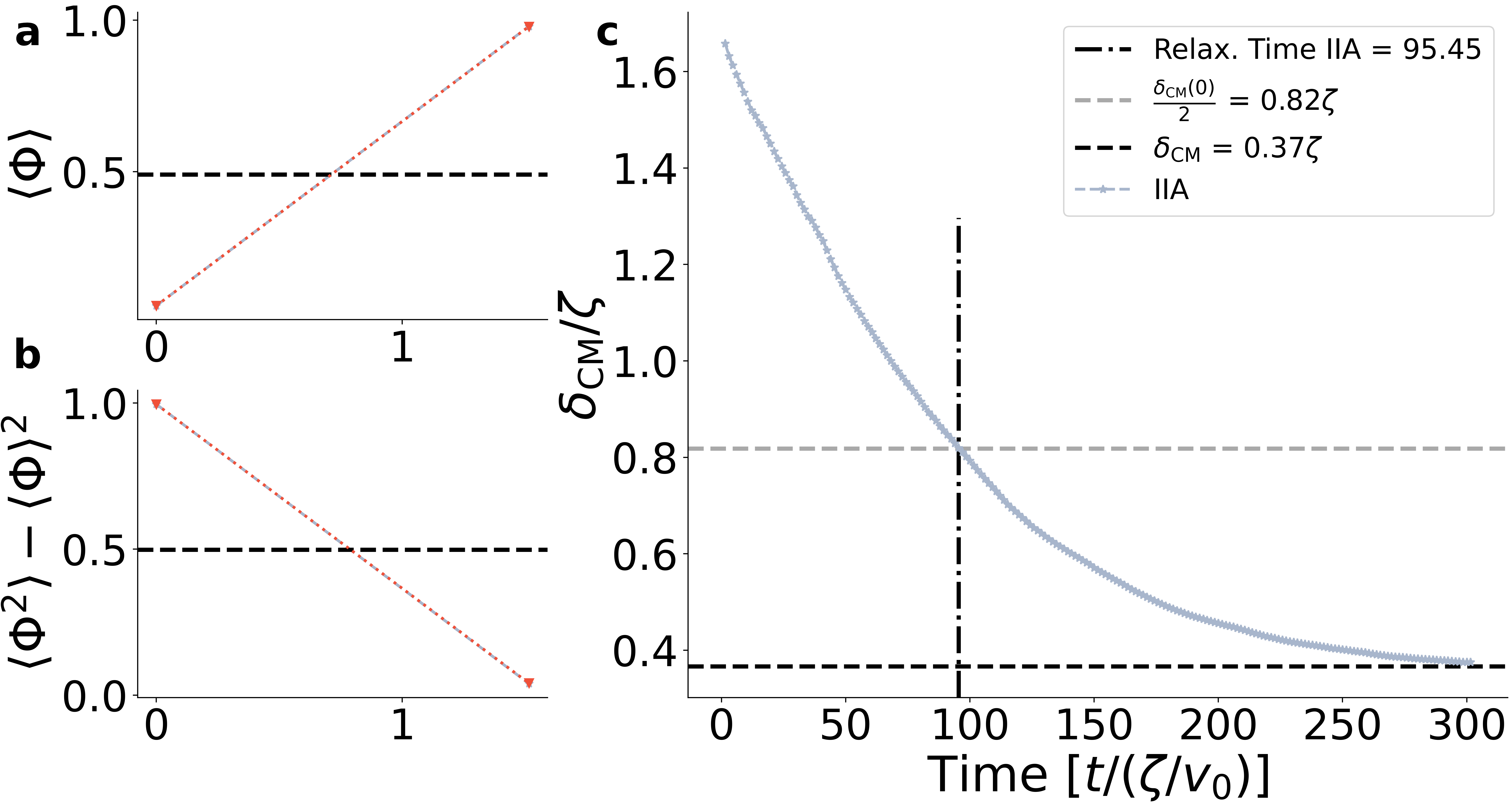}
\caption{\textbf{Relaxation time of the order parameters for the predictive model IIA.} (\textbf{a}-\textbf{c}) For the predictive model IIA, the relaxation times for \textbf{a} the polarization, $\langle\Phi\rangle$, and \textbf{b} the polarization variance, $\langle\Phi^2\rangle-\langle\Phi\rangle^2$ are shorter than $\zeta / v_{0}$. \textbf{c} For the system size, $\delta_{\rm CM}$ drops to half of its value at 95.45$\zeta / v_{0}$. This relaxation times correspond to the parameters as used for Fig.2 in the main text.
}
\label{fig:SI_fig3}
\end{minipage}
\end{figure*}

In Fig.~\ref{fig:SI_fig3} we show the value for the system size, defined as the point when $\delta_{\rm CM}$ drops to half of its initial value, for the predictive model IIA with noise intensities $\eta = 0.1$ and reduced speed $v_{0}/\zeta = 0.0076$ (see Fig.2 in the main text). Following the same definitions, we show the relaxation times for the polarization, $\langle\Phi\rangle$, and the polarization variance, $\langle\Phi^2\rangle-\langle\Phi\rangle^2$.

Finally, the full system evolutions corresponding to the snapshots in Fig.~\ref{fig:SI_fig1} are shown in the supplementary videos SV\_1 (zero noise) and SV\_2 ($\eta = 0.1$).

\section{Minimalist rule for the predictive model}

In the main text, the predictive models IIA and IIB can reproduce flocking behavior for the set of reorientation angles $\Omega_{\theta} = \pm\{0.0, 0.01, 0.2, 0.5\}$. While this relatively broad set of reorientation angles was motivated by the fact that the original Vicsek model allows for arbitrary reorientation, for computational and theoretical reasons, it is reasonable to consider a more minimalistic model with just three allowed reorientations, $\Omega_{\theta} = \pm \{0, \theta'\}$.  

\begin{figure*}
\centering
\begin{minipage}[b]{7in}
\includegraphics[width=0.9 \textwidth,center]{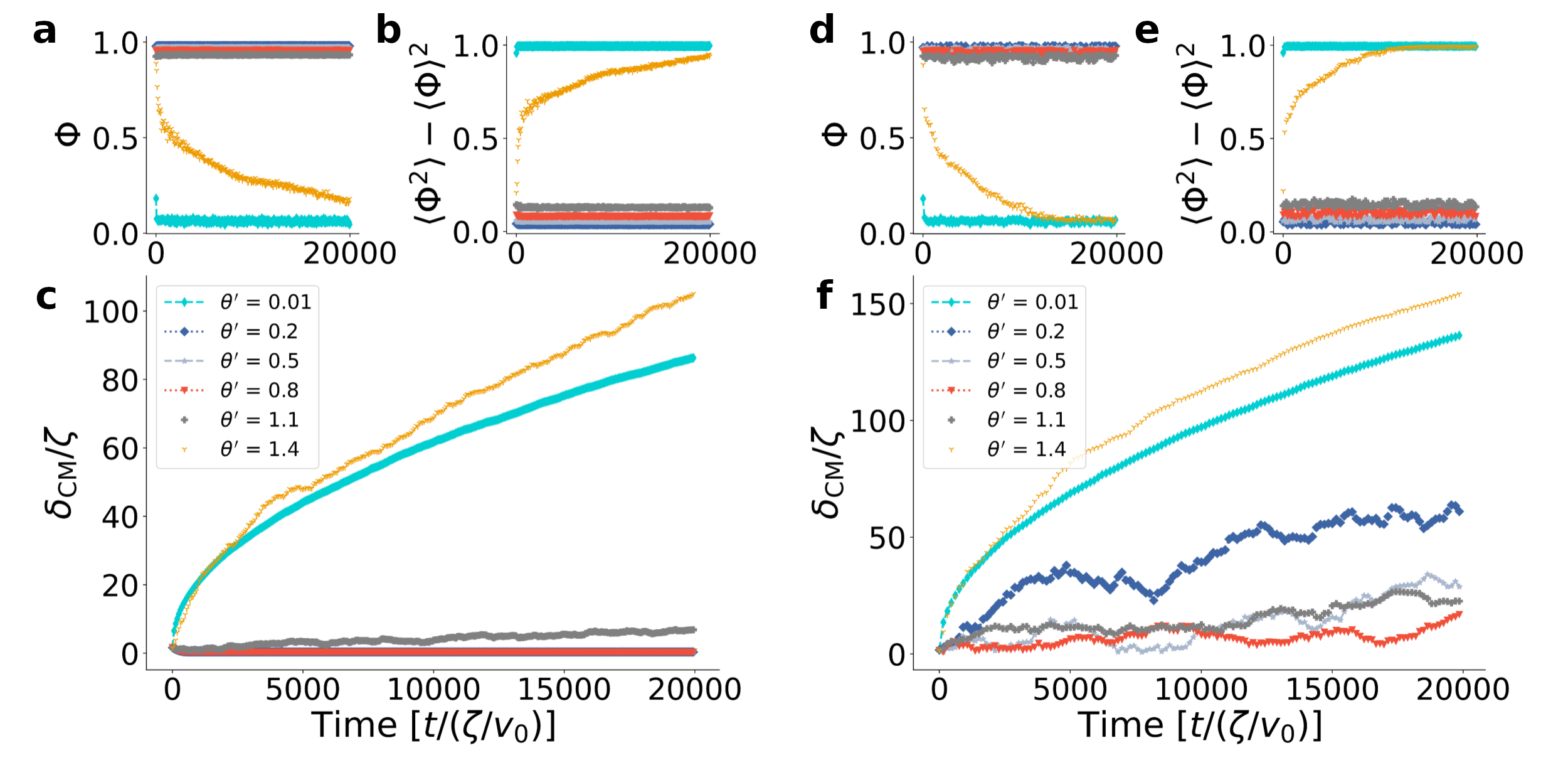}
\caption{\textbf{Relaxation of order parameters for the minimalist version of the predictive model IIA.} (\textbf{a}-\textbf{c}) The time evolution of \textbf{a} average polarization, $\langle\Phi\rangle$, \textbf{b} polarization variance, $\langle\Phi^2\rangle-\langle\Phi\rangle^2$, and \textbf{c} average agent-to-center-of-mass distance, $\delta_{\rm CM}$ for six values of the reorientation angle $\theta'$, noise intensity $\eta=0.1$ agent reduced speed $v_{0}/\zeta = 0.003$. (\textbf{d}-\textbf{f}) The same for a larger reduced speed $v_{0}/\zeta = 0.0076$. Note that the most stable configurations are obtained for $\theta' \approx \pi/4$. The data were averaged over 25 replicas with different noise realizations.
}
\label{fig:SI_fig4}
\end{minipage}
\end{figure*}

\begin{figure*}
\begin{minipage}[b]{7in}
\centering  
\includegraphics[height=0.4
\textwidth,center]{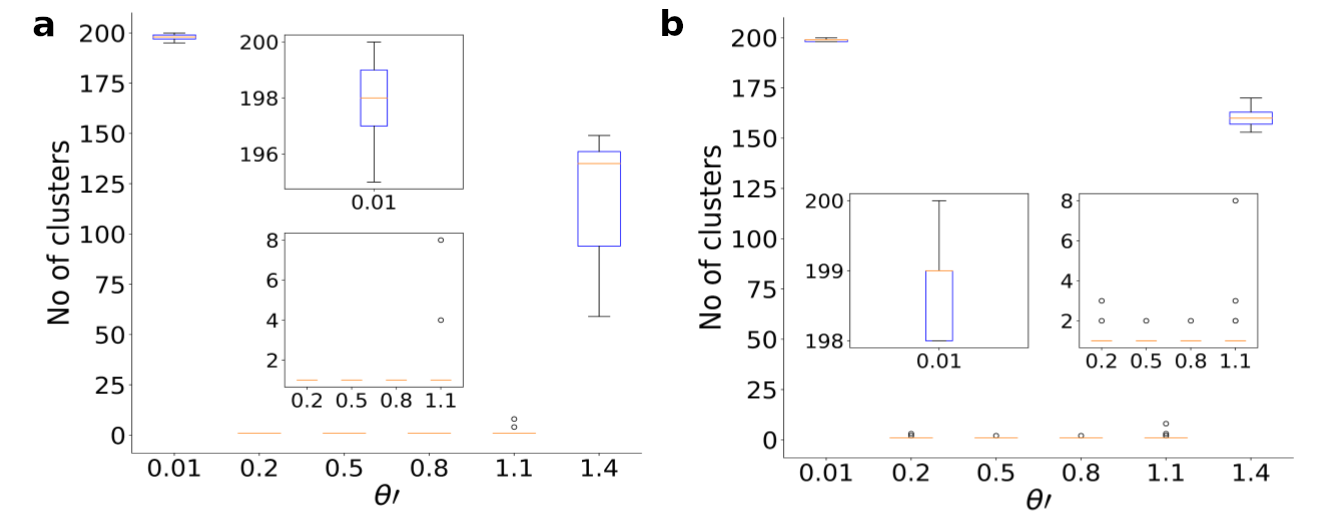}
\caption{\textbf{Number of clusters for the minimalist version of the predictive model IIA.} \textbf{a} Boxplot of the number of communicating clusters at time $t = 2 \times 10^4 \zeta / v_0$ for the simulations with the individual reorientation angles used in Fig.~\ref{fig:SI_fig4}a. The boxplot represents results from 25 independent simulations with different noise realizations. Orange lines indicate the median, boxes span the interquartile range, whiskers extend to data points within 1.5 times the interquartile range, and outliers are shown as individual circles. \textbf{b} The same corresponding to Fig.~\ref{fig:SI_fig4}b.
}
\label{fig:SI_fig5}
\end{minipage}
\end{figure*}

In Figs.~\ref{fig:SI_fig4} and \ref{fig:SI_fig5}, we show that such a minimalistic model indeed yields qualitatively the same behavior as the model investigated in the main text if the distance traveled per timestep over the interaction radius, $v_0\Delta t/\zeta$, is small enough and the angle $\theta'$ is suitably chosen. Our analysis shows that the reorientation angle, which yields the most stable flocks, is $\theta' \approx \pi/4$.  

In Fig.~\ref{fig:SI_fig4}, we show the time evolution of average polarization, $\langle\Phi\rangle$, polarization variance, $\langle\Phi^2\rangle-\langle\Phi\rangle^2$, and average agent-to-center-of-mass distance, $\delta_{\rm CM}$ for $v_0/\zeta = 0.003$ and $0.0076$. In Fig.~\ref{fig:SI_fig5}, we show the corresponding number of communicating clusters at the final simulation time $t = 2 \times 10^4 \zeta / v_0$. In both figures, it can be observed that flock stability increases with decreasing velocity and that $\theta'$ converges to $\pi/4$.

\section{Behavior of the predictive model IIA starting from a polarized state}

\begin{figure*}
\begin{minipage}[b]{7in}
\centering  
\includegraphics[width=1.05
\textwidth,center]{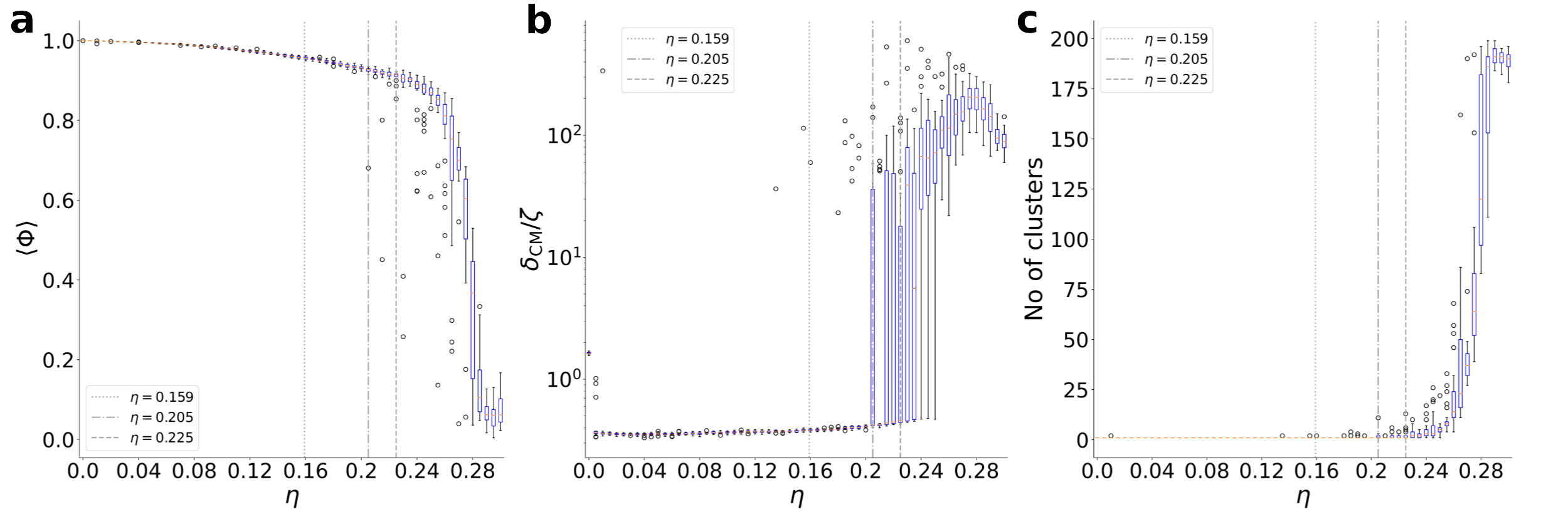}
\caption{\textbf{Effects of the noise for the predictive model IIA, starting from a highly polarized state.} Under the same conditions as in Fig.~3 of the main text but with initially perfectly polarized agents, \textbf{a} the average polarization, \textbf{b} the average agent-to-center-of-mass distance, and \textbf{c} the number of communication clusters exhibit quantitatively the same behavior as in Fig.~3, which corresponds to agents with initially random orientations. In particular, the dynamic transitions marked by the noise values $\eta = 0.159$, $\eta = 0.205$, and $\eta = 0.225$ remain unchanged, indicating their robustness. We simulated 200 particles with the reduced speed of $v_0/\zeta = 0.0076$. The agents start in a perfectly aligned state, with positions uniformly distributed within a box of size $4\zeta \times 4\zeta$. Boxplots represent results from 25 independent simulations with different noise realizations in the way described in the caption of Fig.~\ref{fig:SI_fig5}.
}
\label{fig:SI_fig6}
\end{minipage}
\end{figure*}

Most of the simulation results presented in the main text were obtained for systems initialized with randomly oriented agents. To test the robustness of these results, we show in Fig.~\ref{fig:SI_fig6} that essentially the same outcomes are obtained when the agents are initially aligned. The only difference between the phenomenology described in Fig.3 of the main text (for random initial orientations) and Fig.~\ref{fig:SI_fig6} is that, when the agents are aligned at the start of the simulations, the instability observed for noise intensities $\eta \leq 0.05$, caused by the discrete reorientation angles—which prevent perfect alignment at low noise levels—disappears. This indicates that the dynamic phases discussed in the main text are not a mere artifact of the initial conditions, but a robust feature of the model.

\section{Scalability of the results}

\begin{figure*}
\begin{minipage}[b]{7in}
\centering  
\includegraphics[width=1.05
\textwidth,center]{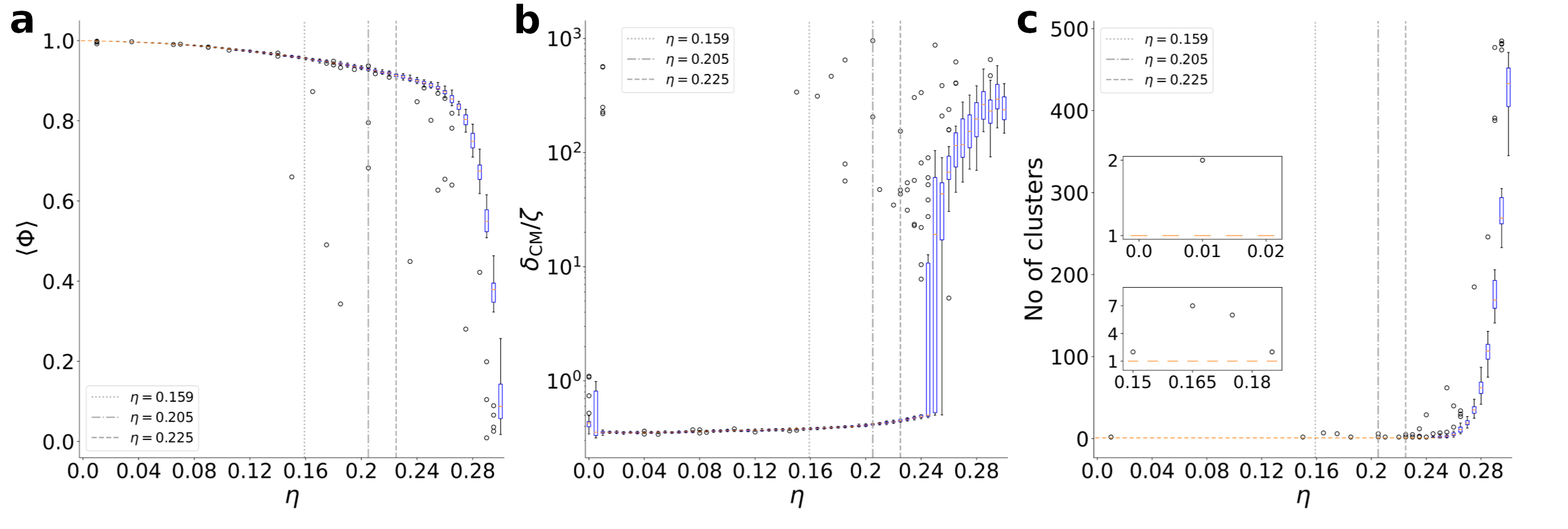}
\caption{\textbf{Effects of noise in the predictive model IIA with $N = 500$ agents.} Under the same conditions as in Fig.~3, we observe that \textbf{a} the average polarization, \textbf{b} the average agent-to-center-of-mass distance, and \textbf{c} the number of communication clusters exhibit qualitatively the same behavior as for $N = 200$. 
The dynamic transition marked by the noise value $\eta = 0.159$, where maximum reorientations due to alignment interactions and noise are equal, and beyond which the number of outliers rapidly increases, remains unchanged. However, the transitions at which the unstable replicas no longer behave as outliers and when most replicas become unstable, previously located at $\eta \approx 0.205$ and $\eta \approx 0.225$ for $N=200$, are shifted by approximately $0.04$ toward higher noise values for $N=500$. This shift indicates that the system with more agents is more stable. Furthermore, the stationary system size
}
\label{fig:SI_fig7}
\end{minipage}
\end{figure*}

To verify the scalability of the presented results, we reproduce in Fig.~\ref{fig:SI_fig7} the analysis from Fig.~3 in the main text but with $N = 500$ agents.  
All qualitative features observed in Fig.~3 are also present for this larger particle number. However, Fig.~\ref{fig:SI_fig7} demonstrates that increasing the particle number enhances system stability. This is reflected, on the one hand, by a reduction in the number of outliers, and more importantly, on the other hand, by a shift of the transition from the stable to the unstable regime toward higher noise values.

\section{Leadership}
\begin{figure*}[ht!]
\centering
\begin{minipage}[b]{7in}
    \includegraphics[width=0.9
\textwidth,center]{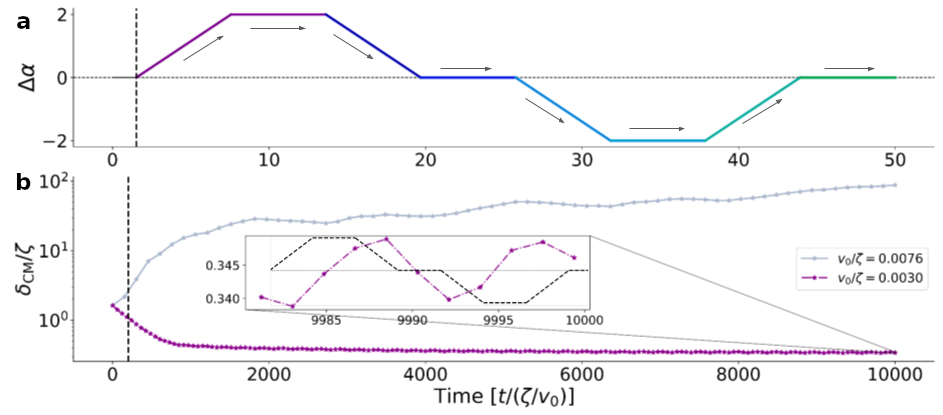}
\caption{\textbf{Leadership in the predictive model IIA}  
In this numerical experiment, the agents were initially perfectly aligned. After an equilibration period of 200 timesteps, a subgroup of 20 leaders was selected to change orientation deterministically according to the oscillatory protocol shown in \textbf{a}, with $v_0/\zeta = 0.0076$. The leaders reoriented with an angular velocity of $\Delta \alpha = 0.0025$~rad/timestep for 800 timesteps, interspersed with relaxation periods of 800 timesteps.  
\textbf{b} Under this protocol, for $v_0/\zeta = 0.0076$ and noise intensity $\eta = 0.04$, the flock disperses, as indicated by the increase in the average agent-to-center-of-mass distance, $\delta_{\rm CM}$. In contrast, stable flocking is maintained in the absence of leaders. However, when the reduced speed is decreased to \( v_0/\zeta = 0.003 \), the agents successfully follow the leaders, forming a characteristic pattern where maxima in system size lag behind the leaders' turning events.
 This behavior is highlighted in the inset, which magnifies a single oscillation of $\delta_{\rm CM}$ at the time marked by the vertical dashed line (pink dash-dotted line). As a visual reference, the inset also includes the corresponding angular variation of the leaders from \textbf{a} (black dashed line).  
}
    \label{fig:fig5}
\end{minipage}
\end{figure*}

Figure~\ref{fig:fig5} shows that the predictive model IIA can form cohesive flocks also in the scenario when a subgroup of leaders performs an oscillator deterministic motion, albeit for slightly lower $v_0/\zeta$ than without the perturbation by leaders. The video SV\_3 shows the stationary flock following the leaders.

\section{Supplementary videos}

\subsection*{Supplementary video 1: Relaxation of the system for a vanishing noise}

SV\_1: 
Behavior of one of the replicas of the predictive model IIA for a system of 200 particles with noise $\eta = 0.0$ and reduced speed $v_0/\zeta = 0.0076$. The agents start with uniformly distributed orientations in $[0,2\pi]$ and positions within a box of size $4\zeta \times 4\zeta$. The total simulation time is $20000 \zeta / v_0$. After a transient period of duration $12100 \zeta / v_0$, the system reaches a stationary V-shaped configuration. During the relaxation process, as well as in the stationary state, the agents constantly change positions, even though the simulation runs in the absence of noise

\subsection*{Supplementary video 2: Relaxation of the system for a finite noise}
SV\_2: Behavior of one of the replicas of the predictive model IIA for a system of 200 particles with noise $\eta = 0.1$ and reduced speed $v_0/\zeta = 0.0076$. The agents start with uniformly distributed orientations in $[0,2\pi]$ and positions within a box of size $4\zeta \times 4\zeta$. The simulation time is $20000 \zeta / v_0$. After a transient period of duration $251 \zeta / v_0$, the system reaches a stationary circular configuration.

\subsection*{Supplementary video 3: Leadership}

SV\_3: Behavior of one of the replicas for a system of 200 particles with noise $\eta = 0.04$ and reduced speed $v_0/\zeta = 0.0076$, using the leadership protocol with 20 leaders. The agents start with perfectly aligned orientations and uniformly distributed positions within a box of size $4\zeta \times 4\zeta$. The simulation time is $20000 \zeta / v_0$. After an initial equilibration period of 200 timesteps, the leaders reorient deterministically with an angular velocity of $\Delta \alpha = 0.0025$~rad/timestep for 800 timesteps, interspersed with relaxation periods of 800 timesteps. The other agents obey the predictive model IIA.

\clearpage

\bibliography{Main}